\documentclass[reprint,11pt]{elsarticle}

\usepackage{lineno,hyperref}
\usepackage{amsmath}
\usepackage{amssymb}
\usepackage{mathtools}

\usepackage{fullpage}
\usepackage{todonotes}

\newcommand{\nua}[1]{\ensuremath{\rlap{\kern-2.5pt\ensuremath{\overset{\scriptscriptstyle(-)}{\phantom{\nu}}}}{\ensuremath{{\nu}_{#1}}}}}

\journal{Physics Letters B}

\begin{document}

\begin{frontmatter}

\title{The gallium anomaly revisited}

\author{J. Kostensalo, J. Suhonen, }
\address{University of Jyvaskyla, Department of Physics, P.O. Box 35, FI-40014, Finland}
\author{C. Giunti}
\address{INFN, Sezione di Torino, Via P. Giuria 1, I-10125 Torino, Italy}
\author{P. C. Srivastava}
\address{Department of Physics, Indian Institute of Technology, Roorkee 247667, India}

\begin{abstract}
The gallium anomaly, i.e. the missing electron-neutrino flux from $^{37}$Ar and
$^{51}$Cr electron-capture decays as measured by the GALLEX and
SAGE solar-neutrino detectors, has been among us already for about two decades. 
We present here a new estimate of the significance of this anomaly based on 
cross-section calculations using nuclear shell-model wave functions obtained 
by exploiting recently developed two-nucleon interactions. The gallium anomaly of the 
GALLEX and SAGE experiments
is found to be smaller than that obtained in previous evaluations, decreasing the 
significance from 3.0$\sigma$ to 2.3$\sigma$. 
This result is compatible with the recent indication in favor of short-baseline
$\bar\nu_{e}$ disappearance due to small active-sterile neutrino mixing
obtained from the combined analysis of the data of the NEOS and DANSS reactor experiments.
\end{abstract}


\end{frontmatter}


\section{Introduction}

Gallium-based solar-neutrino experiments, 
GALLEX \cite{Anselmann1995,Hampel1998,Kaether2010} and 
SAGE \cite{Abdurashitov}, were designed to detect $pp$ neutrinos from the sun.
These two experiments are unique in having been tested for their detection
efficiency by $^{37}$Ar and $^{51}$Cr radioactive sources. These sources 
emit discrete-energy electron neutrinos ($E_{\nu}<1 $ MeV) based on their decay 
via nuclear electron capture (EC). Detection of these neutrinos is achieved 
through the charged-current neutrino-nucleus scattering reaction
\begin{equation}
\nu_e + \,^{71}{\rm Ga}(3/2^-)_{\rm g.s.} \ \rightarrow \, ^{71}{\rm Ge}(J^{\pi}) + e^- 
\label{eq:CC-71}
\end{equation}
to the lowest four (flux from the $^{51}$Cr source) or five (flux from the $^{37}$Ar 
source) nuclear states in $^{71}\rm Ge$. 
In this article we discuss also the analogous reaction
\begin{equation}
\nu_e + \,^{69}{\rm Ga}(3/2^-)_{\rm g.s.} \ \rightarrow \, ^{69}{\rm Ge}(J^{\pi}) + e^-
\label{eq:CC-69}
\end{equation}
in order to test our calculated shell-model wave functions more comprehensively.

The scattering of
$^{37}$Ar and $^{51}$Cr neutrinos off $^{71}$Ga leads mainly to the ground state
and the excited states at 175 keV and 500 keV in $^{71}$Ge. 
The scattering cross sections for
the mentioned three low-lying states can be estimated by using the data from
charge-exchange reactions \cite{Frekers:2011zz}
or by using a microscopic nuclear model, 
like the nuclear shell model \cite{Bahcall1997}. In both cases it has been found that the
estimated cross sections are larger than the ones measured by the 
GALLEX \cite{Anselmann1995,Hampel1998,Kaether2010} and 
SAGE \cite{Abdurashitov} experiments. The measured neutrino capture rates (cross 
sections) are 0.87 $\pm$ 0.05 of the rates based on the 
cross-section estimates by Bahcall \cite{Bahcall1997}. The related  model
calculations and analyses based on them have been discussed in 
\cite{Giunti2011,Haxton:1998uc,Giunti:2012tn}. It should be noted that the response to the 
ground state is known from the electron-capture $ft$ value of $^{71}$Ge.  
The discrepancy between the measured and theoretical capture rates constitutes
the so-called ``gallium anomaly''.

One of the explanations to the the gallium anomaly is associated with the oscillation 
of the electron neutrinos to a sterile neutrino in
eV mass scale \cite{Giunti2011,Giunti:2012tn}. The same scheme
could also explain the so-called ``reactor-antineutrino anomaly'' 
\cite{Mueller2011,Mention:2011rk,Huber:2011wv}, discussed, e.g. in \cite{Giunti:2012tn}. 
Searches for the sterile neutrinos 
are under progress in several laboratories. However, it should be remarked here 
that there is no accepted sterile neutrino model to explain the experimental 
anomalies consistently, and also alternative solutions to the
reactor-antineutrino anomaly have been proposed, like the proper
inclusion of first-forbidden $\beta$-decay branches in the construction of the
cumulative antineutrino spectra \cite{Hayen2019}.

\section{Neutrino-nucleus scattering formalism}

We now summarize the main points of the formalism for calculating cross sections 
for charged-current neutrino-nucleus scattering. Details of the formalism can be 
found from \cite{Ydrefors2012,Walecka2004}.

For the low-energy ($E_{\nu}<1 $ MeV) $^{37}\rm Ar$ and $^{51}\rm Cr$ neutrinos 
considered in this work the creation of the two heavier lepton 
flavors, $\mu$ and $\tau$, is not energetically possible.  At these low energies 
the four-momentum transfer is small compared to the mass 
of the exchanged gauge boson $W^{\pm}$, that is, $Q^2=-q_{\mu}q^{\mu} \ll M^2_{W^{\pm}}$.
Therefore, to a good approximation the scattering can be considered in the lowest 
order as a single effective vertex with a coupling constant 
$G=G_{\rm F}\cos(\theta_{\rm C})$, where $G_{\rm F}$ is the 
Fermi constant and $\theta_{\rm C}\approx 13 ^{\circ} $ is the Cabibbo angle. The matrix 
element of this effective Hamiltonian can be written as
\begin{equation}
\langle f\vert H_{\rm eff} \vert i\rangle = \frac{G}{\sqrt{2}} \int d^3\mathbf{r}l_{\mu}
e^{-\mathbf{q}\cdot \mathbf{r}}\langle f\vert \mathcal{J}^{\mu}(\mathbf{r})\vert i\rangle \;,
\label{eq:effme}
\end{equation}
where $\mathcal{J}^{\mu}$ denotes the hadronic current and 
$l_{\rm \mu}=e^{\mathbf{q}\cdot \mathbf{r}}\langle l\vert j_{\mu}(r)\vert\nu\rangle$ 
\cite{Ydrefors2012}. 

The initial nuclear state in the scatterings of Eqs.~(\ref{eq:CC-71}) and 
(\ref{eq:CC-69}) is the $J_i^{\pi_i}=3/2^-$ ground state of $^{69,71}\rm Ga$. 
Assuming that the final 
nuclear states in $^{69,71}\rm Ge$ also have well defined 
spin-parities $J_f^{\pi_f}$, the double differential cross section for the 
charged-current (CC) neutrino-nucleus scattering is given 
by \cite{Ydrefors2012,Kolbe2003,Ydrefors2013}
\begin{align}
\notag
\left[ \frac{d^2 \sigma_{i\rightarrow f}}{d\Omega d E_{\rm exc}} \right] &= 
\frac{G^2|\mathbf{k}_{l}|E_{l}}{\pi(2J_i+1)}F(Z_f,E_{l}) \\
  & \times \left( \sum_{J\geq 0} \sigma_{\rm CL}^J +  \sum_{J\geq 1} \sigma_{\rm T}^J  \right),
\label{eq:ddcs}
\end{align}
where $\mathbf{k}_{l}$ and $E_{l}$ are the three-momentum and energy of the 
outgoing lepton, respectively, and $F(Z_f,E_{l})$ is the Fermi function 
which accounts for the Coulomb interaction of the low-energy final-state lepton 
and the residual nucleus \cite{Engel1998}. 
Here $\sigma_{\rm CL}^J$ is the Coulomb-longitudinal component  
and $\sigma_{\rm T}^J$ is the transverse component. Detailed formulas for these can 
be found in Ref. \cite{Walecka2004}. The operators contain vector and axial-vector 
pieces, which depend on the four-momentum-transfer-dependent nuclear form 
factors $F^{\rm V}_{1,2}$ (vector), $F^{\rm A}$ (axial-vector), and $F^{\rm P}$ 
(pseudoscalar). At low neutrino energies the cross section is dominated by 
Fermi and Gamow-Teller type of transitions which proceed via the operators 
$F^{\rm V}(q)j_0(qr)\textbf{1}$ and $F^{\rm A}(q)j_0(qr)\boldsymbol{\sigma}$ 
respectively \cite{Ydrefors2012}. There are also small contributions from 
spin-dipole type transitions mediated by the operator 
$F^{\rm A}(q)[j_1(qr)\boldsymbol{Y}_1\boldsymbol{\sigma}]_{0^-,1^-,2^-}$. 

\section{Results of nuclear-structure calculations}

\begin{figure}[htb]
	\centering
	\includegraphics[width=0.60\textwidth]{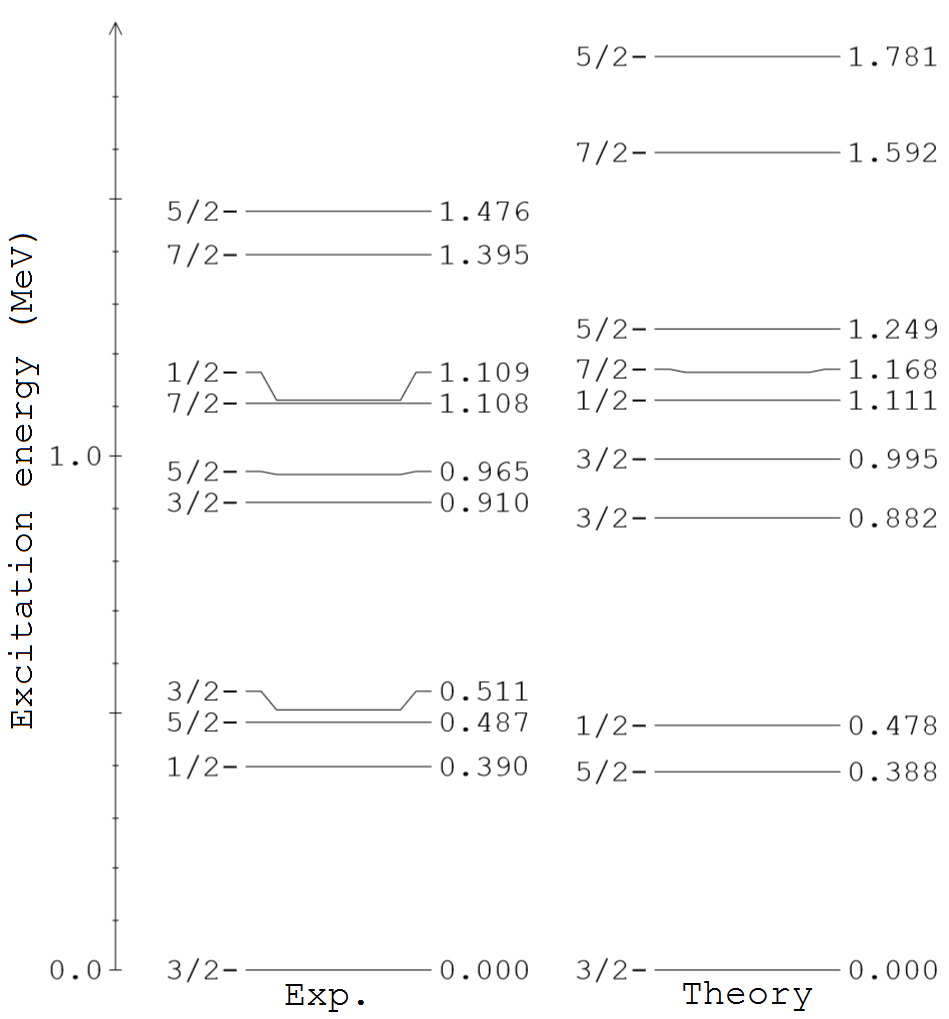}
	\caption{Experimental and theoretical low-lying energy spectra of $^{71}\rm Ga$. 
\label{fig:Gaspec}}
	\end{figure}

\begin{figure}[htb]
	\centering
	\includegraphics[width=0.60\textwidth]{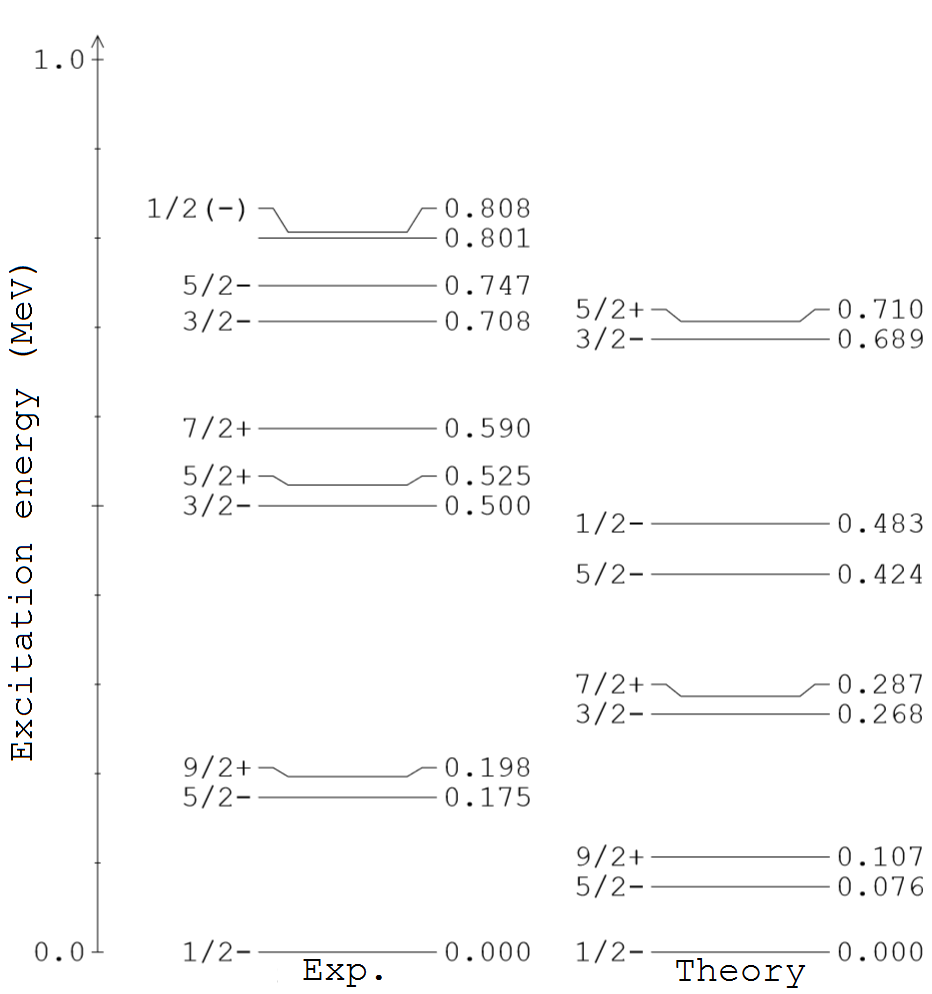}
	\caption{Experimental and theoretical low-lying energy spectra of 
$^{71}\rm Ge$.\label{fig:Gespec}}
	\end{figure}

The nuclear wave functions and one-body transition densities (OBTDs) 
(see e.g. \cite{Suhonen2007}) were calculated 
in the interacting nuclear shell model using the computer code 
NuShellX@MSU \cite{nushellx}. The calculations were done in a model space 
consisting of the proton and neutron orbitals $0f_{5/2}$, $1p$, and $0g_{9/2}$ with 
the effective Hamiltonian JUN45 \cite{Honma2009}. The low-energy excitation 
spectra of $^{71}\rm Ga$ and $^{71}\rm Ge$, of interest in this work, are 
presented in Figs.~\ref{fig:Gaspec} and \ref{fig:Gespec}, respectively 
(see also Honma et al. \cite{Honma2009} for the 1--3 MeV range in $^{71}\rm Ge$). 
For both cases the ground-state spin-parity is correctly predicted: 
$3/2^-$ for $^{71}\rm Ga$ and $1/2^-$ for $^{71}\rm Ge$. The energies of the first 
two excited states in $^{71}\rm Ga$ agree well with the experimental spectrum but 
the ordering of the $5/2^-$ and $1/2^-$ states is reversed. The second $3/2^-$ 
state is also higher than the experimental one, with energy 882 keV compared to 
the experimental energy 511 keV. For $^{71}\rm Ge$ the ordering of the first four 
states, including the negative parity states which we are actually interested in, 
agree with the experimental data. The qualitative features of the computed low-energy 
spectrum are also very similar to the experimental one, with the gap between 
the $5/2^-$ and $9/2^+$ states being much narrower than the gap between the 
ground state and the first exited state as well as the gap between the 
$9/2^+$ and $3/2^-$ states. However, the shell-model-calculated energies of the 
excited states are a bit lower than the experimentally determined states, with 
the 175 keV state predicted at 76 keV and the 500 keV state predicted at 268 keV. 

The electromagnetic properties are also pretty well predicted (see the original 
article of Honma et al. \cite{Honma2009} for details). For the ground state 
of $^{71}\rm Ga$ the theoretical electric quadrupole moment 
is +0.155 $eb$ \cite{Honma2009} (with effective charges $e_p=1.5,e_n=0.5$) 
while the experimental one is +0.1040(8) $eb$ \cite{Pernpointner1998}. 
For the $9/2^+$ state in $^{71}\rm Ge$ the theoretical value -0.339 $eb$ seems to 
agree perfectly with the experimental value 0.34(5) $eb$ \cite{Bertschat1970}, 
however, the sign has not been experimentally determined. The magnetic dipole 
moments of the low-lying states are excellently predicted. With the $g$ factors 
$g_l=g_l^{\rm (free)}$ and $g_s=0.7g_s^{\rm (free)}$ the magnetic dipole moment of 
the $^{71}\rm Ga$ ground state is predicted to be +2.188 $\mu_N$ which is in good 
agreement with the experimental value +2.56227(2) $\mu_N$ \cite{Pernpointner1998}. 
For the $1/2^-$, $5/2^-$, and $9/2^+$ states in $^{71}\rm Ge$ the magnetic moments 
are +0.438 $\mu_N$, +1.060 $\mu_N$, $-$1.014 $\mu_N$, while the experimental 
ones are +0.547(5) $\mu_N$ \cite{Childs1966}, +1.018(10) 
$\mu_N$ \cite{Morgenstern1968}, and $-$1.0413(7) $\mu_N$ \cite{Bertschat1970},
respectively.

The calculations were also done for $^{69}\rm Ge$ and $^{69}\rm Ga$. In this case 
the shell model reproduced the experimental data even better. The first three 
states in $^{69}\rm Ge$ were predicted with the correct spin parities $5/2^-$, 
$1/2^-$, and $3/2^-$. The shell model energies 0 keV, 98 keV, and 189 keV agree 
well with the experimental energies 0 keV, 87 keV and 233 keV. Also the 
electromagnetic properties are well reproduced in this case \cite{Honma2009}.

For the calculation of the cross sections we use recently measured $Q$ values 
and branching ratios from \cite{nndc} and 
the same $L/K$ capture ratios as well as atomic overlap corrections as Bahcall 
in his analysis \cite{Bahcall1997}. The neutrino energies adopted here 
are for $^{51}\rm Cr$ 751.82 keV (9.37 \%), 746.99 keV (80.70 \%), 431.74 keV  
(1.033 \%), and 426.91 keV (8.890 \%). For $^{37}\rm Ar$ the energies we use are
813.60 keV (90.2 \%) and 811.05 keV (9.8 \%). 

\section{Results for BGT values}

The neutrino-nucleus scattering cross sections are proportional to the $\beta$-decay 
BGT values, which could be, in principle, extracted from $\beta$-decay half-lives. 
This procedure gives us accurately the ground-state-to-ground-state BGT value, 
but it is not implementable for the BGTs of the excited states. Therefore, 
other techniques, such as performing charge-exchange reactions, must be utilized. 
However, this technique can be problematic for some transitions due to the 
significant tensor contributions, as was shown to be the case for the $(p,n)$ reaction
leading to the first excited state in $^{71}\rm Ga$ by Haxton \cite{Haxton:1998uc}. 
In this case, there is a significant cancellation between the Gamow-Teller (GT) 
and tensor (T) matrix elements. The interference between the GT and T NMEs
is described by the linear combination
\begin{equation}
\langle f || O_{(p,n)} || i \rangle  = \langle f || O_{\rm GT} || i \rangle 
+\delta \langle f || O_{L=2} || i \rangle  \;,
\label{eq:GT-T}
\end{equation}
where $i$ ($f$) is the initial (final) nuclear state and $\delta$ is the mixing 
parameter.

		\begin{table}[htb]
\caption{Results for $^{71}\rm Ga$ with $\delta = 0.097$ in Eq.~(\ref{eq:GT-T}).}
		\centering
\begin{tabular}{lllll}
\hline
State & $\langle f || O_{\rm GT} || i \rangle $ &  $\langle f || O_{L=2} || i \rangle $  
& $\rm BGT_{\beta}^{SM}$ & $\rm BGT^{SM}_{(p,n)}$ \\
\hline
$1/2^-_{\rm g.s.}$& -0.795 & 0.465 & 0.158 & 0.141 \\
$5/2^-_1$& 0.144 & -1.902 & 0.0052 & 0.0004\\
$3/2^-_1$ & 0.100 & 0.0482 & 0.0025 & 0.0027 \\
$3/2^-_2$ & 0.430 & -0.0014 & 0.0462 & 0.0462 \\
$1/2^-_2$ & -0.620 & 0.348 & 0.0958 & 0.0857 \\ 
\hline
\end{tabular}
\label{tbl:71me}
\end{table}

		\begin{table}[htb]
  \caption{ The results for $^{69}\rm Ga$ with $\delta = 0.097$ in Eq.~(\ref{eq:GT-T}).} 
		\centering
\begin{tabular}{lllll}
\hline
State & $\langle f || O_{\rm GT} || i \rangle $ &  $\langle f || O_{L=2} || i \rangle $  
& $\rm BGT_{\beta}^{SM}$ & $\rm BGT^{SM}_{(p,n)}$ \\
\hline
$5/2^-_{\rm g.s.}$& -0.0139 & -1.180 & $4.802\times 10^{-5}$ & $4.117 \times 10^{-3}$  \\
$1/2^-_1$& -0.592 & 0.238 & 0.0876 & 0.0809 \\
$3/2^-_1$ & 0.0298 & 0.422 & $2.220\times 10^{-4}$ & $1.251 \times 10^{-3}$ \\
\hline
\end{tabular}
\label{tbl:69me}
\end{table}

The calculated GT and T matrix elements are listed in tables \ref{tbl:71me} 
and \ref{tbl:69me}. For $^{71}\rm Ga$ the matrix elements for the scattering to 
the $3/2^-_2$ and $1/2^-_2$ states are also included for comparison with the 
available charge-exchange data. As predicted by Haxton \cite{Haxton:1998uc}, 
the GT and T contributions cancel significantly for $5/2^-_1$. For the $3/2^-_1$ 
state the contributions are constructive. Interestingly, the GT and T contributions 
counteract each other also for the ground-state-to-ground-state transition, 
meaning that a $(p,n)$ reaction would underestimate the BGT value. This is 
significant regarding the validity of the BGT values reported by 
Frekers et al. \cite{Frekers:2011zz,Frekers2015}, since there these tensor 
contributions are ignored. This leads to an 
underestimation of the ground-state-to-ground-state BGT value and, since this 
is adjusted to the one extracted from $\beta$ decay, to an overestimation of the 
BGT values to the excited states. It should be noted that the value 
$\delta=0.097$ used here, as well as in the work of Haxton, has been obtained 
by fitting $\beta$ transitions in the $p$-shell \cite{Haxton:1998uc}. However, 
uncertainties related to this choice are hard to quantify. A reasonable estimate 
might be  $0.05 <\delta <0.15$, meaning that the overestimation in the ground-state
BGT is somewhere between 10 \% and 40 \%. It should be emphasized that this 
alone is not enough to explain the discrepancy between the shell-model 
calculations and charge-exchange results, since the ratio 
$\rm BGT_{500}/BGT_{g.s.}$ is $0.207 \pm 0.016$ according to the charge-exchange 
experiment \cite{Frekers2015}, while the shell model predicts a ratio 
as low as 0.019. 

In \cite{Frekers:2011zz} the projectile, target and relative angular-momentum transfers
$[J_{\rm projectile}\  J_{\rm target} \ J_{\rm relative}]$ were measured in the
$^{71}$Ga($^3$He,$^3$H)$^{71}$Ge charge-exchange reaction. One possible explanation for 
the remaining difference between the shell-model calculations and charge-exchange 
results relates to the extraction of the $[110]$ component of the
angular-momentum transfers at $0^{\circ}$, which 
corresponds to the GT and T contributions. This was done in \cite{Frekers:2011zz} 
by fitting various angular-distribution functions, with different 
$[J_{\rm projectile}\  J_{\rm target} \ J_{\rm relative}]$ combinations, to the experimental angular 
distribution. However, in the calculation of the distributions shell-model OBTDs 
calculated in the $fp$-space using the Hamiltonian GXPF1a \cite{gx1,gx2} were used. 
This Hamiltonian does not seem to be the best choice here: it for example predicts 
the level ordering of $^{71}\rm Ge$ as $5/2^-$ ground state, $1/2^-$ at 388 keV, 
and $3/2^-$ at 1496 keV, which does not agree at all with the experimental spectrum. 
The one-body transition densities turn out to be off as well. To replicate the 
experimental half-life of $^{71}\rm Ge$ one would need to adopt $g_{\rm A}\approx 0.6$ 
and the ratios $\rm BGT_{175}/BGT_{g.s.}$ and $\rm BGT_{500}/BGT_{g.s.}$ are predicted 
as 0.0025 and 0.695 respectively, which is not at all consistent with the final 
experimental values. Frekers \textit{et al.} report the [110] component at $0^{\circ}$ to 
be 92 \% for the ground state and 87 \% for the second excited state. It cannot 
be easily estimated how much and which way the use of these OBTDs effects the 
fits and thus the percentages. In a scenario where the ground-state [110] component 
is underestimated and/or the 500 keV-state [110] component is overestimated, 
we would also have an other source of systematic overestimation of the ratio 
$\rm BGT_{500}/BGT_{g.s.}$. For example if the true [110] components for the ground 
state and the $3/2^-$ state would be 95\% and 70\% instead and $\delta=0.15$, 
we would get roughly a 70\% overestimate for the BGT ratio.

What comes to the transitions to the $1/2^-_2$ and $3/2^-_2$ states, the shell 
model is in agreement with the charge-exchange results in that the transition to 
the $1/2^-_2$ state is the second strongest after the ground-state-to-ground-state 
transition. However, the transition to $3/2^-_2$ is predicted to be significantly 
stronger than the one to the $3/2_1$ state, while the results of 
Frekers et al. \cite{Frekers2015} would imply this to be the weakest of the 
transitions. The shell model predicts qualitatively correctly that the 
ground-state-to-ground-state scattering has a much lower cross section for 
$^{69}\rm Ga$ than for $^{71}\rm Ga$, but the ratio $\approx 3\times 10^{-4}$ seems 
to be off from the experimental value $\approx 0.02$. The inability of the 
shell model to predict very low BGT values is due to the fact that there are 
cancellations of single-particle matrix elements of roughly the same size, 
resulting in large numerical inaccuracies. However, this is not a problem for the
larger BGT values where theoretical uncertainties are usually 
about 10 \% \cite{Brown1985}. On the other hand, the BGT values for the excited states 
in $^{71}\rm Ge$ are rather small, but there should not be any problems with the 
numerical inaccuracies as large cancellations are not present. Here we adopt 
a very conservative 50 \% uncertainty for these transitions in order to avoid 
overstatements regarding the significance of the gallium anomaly.    

\section{Results for scattering cross sections}

		\begin{table}[htb]
  \caption{Cross-section results for the $^{51}\rm Cr$ neutrinos with JUN45 interaction. 
The cross sections are in units cm$^2$.} 
		\centering
\begin{tabular}{cc}
\hline
State & $g_{\rm A}=0.955(6)$ \\
\hline
$1/2^-_{\rm g.s.}$& $5.53\pm0.07  \times 10^{-45}$   \\
$5/2^-_1$& $  1.21 \pm  0.61   \times10^{-46}$ \\
$9/2^+_1$ & $\leq 10^{-56}$ \\
$3/2^-_1$ & $  1.94 \pm  0.97   \times10^{-47}$  \\
total & $ 5.67  \pm  0.10  \times 10^{-45}$ \\
\hline 
\label{tbl:751}
\end{tabular}
\end{table}

		\begin{table}[htb]
  \caption{Cross-section results for the $^{37}\rm Ar$ neutrinos with JUN45 interaction. 
The cross sections are in units cm$^2$.} 
		\centering
\begin{tabular}{cc}
\hline
State & $g_{\rm A}=0.955(6)$ \\
\hline
$1/2^-_{\rm g.s.}$& $6.62\pm0.09 \times 10^{-45}$   \\
$5/2^-_1$& $ 1.51  \pm  0.76   \times10^{-46}$ \\
$9/2^+_1$ & $\leq 10^{-56}$ \\
$3/2^-_1$ & $ 2.79  \pm  1.40   \times10^{-47}$  \\
$5/2^+_1$ & $5.91 \pm 2.96 \times 10^{-51}$ \\
total & $ 6.80  \pm 0.12   \times 10^{-45}$  \\
\hline
\label{tbl:814}
\end{tabular}
\end{table}

The cross sections for the $^{51}\rm Cr$ and $^{37}\rm Ar$ neutrinos scattering off 
$^{71}\rm Ga$ are given in tables \ref{tbl:751} and \ref{tbl:814}. The contributions 
of the excited states are about 2.5(1.3) \%. The contributions of the positive-parity 
states are about $10^{-4}$ \% and thus the fact that these were left out from 
the previous analyses does not affect the reliability of their conclusions.

\section{Reassessment of the gallium anomaly}

\begin{table}[!tb]
\caption{
\label{tab:crs}
Gallium cross sections (in units of $10^{-45} \, \text{cm}^2$)
for ${}^{51}\text{Cr}$ and ${}^{37}\text{Ar}$ neutrinos
and their ratios with the central value of the corresponding Bahcall cross section 
\protect\cite{Bahcall:1997eg}
in the first line.
The other lines give the cross sections corresponding to the BGT's of
Haxton
\protect\cite{Haxton:1998uc,Giunti:2012tn},
Frekers et al.
\protect\cite{Frekers:2011zz,Giunti:2012tn},
and the JUN45 calculation presented in this paper.
}
\centering
{
\renewcommand{\arraystretch}{1.45}
\begin{tabular}{lcccc}
\hline
&
$\sigma^{^{51}\text{Cr}}$
&
$\sigma^{^{51}\text{Cr}}/\sigma^{^{51}\text{Cr}}_{\text{B}}$
&
$\sigma^{^{37}\text{Ar}}$
&
$\sigma^{^{37}\text{Ar}}/\sigma^{^{37}\text{Ar}}_{\text{B}}$
\\
\hline
Bahcall
&
$
5.81
\pm
0.16
$
&
&
$
7.00
\pm
0.21
$
&
\\
Haxton
&
$
6.39
\pm
0.65
$
&
$
1.100
\pm
0.112
$
&
$
7.72
\pm
0.81
$
&
$
1.103
\pm
0.116
$
\\
Frekers
&
$
5.92
\pm
0.11
$
&
$
1.019
\pm
0.019
$
&
$
7.15
\pm
0.14
$
&
$
1.021
\pm
0.020
$
\\
JUN45
&
$
5.67
\pm
0.06
$
&
$
0.976
\pm
0.011
$
&
$
6.80
\pm
0.08
$
&
$
0.971
\pm
0.011
$
\\
\hline
\end{tabular}
}
\end{table}

\begin{table}[!h]
\caption{
\label{tab:bgt}
Values of the Gamow-Teller strengths of the transitions
from the ground state of ${}^{71}\text{Ga}$ to the relevant excited states of ${}^{71}\text{Ge}$
relative to the
Gamow-Teller strength of the transitions to the ground state of ${}^{71}\text{Ge}$
obtained by
Krofcheck et al.
\protect\cite{Krofcheck:1985fg,Krofcheck-PhD-1987},
Haxton
\protect\cite{Haxton:1998uc},
Frekers et al.
\protect\cite{Frekers:2011zz},
and with the JUN45 calculation presented in this paper.
}
\centering
\resizebox{\textwidth}{!}
{
\renewcommand{\arraystretch}{1.45}
\begin{tabular}{lcccc}
\hline
&
Method
&
$\dfrac{ \text{BGT}_{5/2-} }{ \text{BGT}_{\text{gs}} }$
&
$\dfrac{ \text{BGT}_{3/2-} }{ \text{BGT}_{\text{gs}} }$
&
$\dfrac{ \text{BGT}_{5/2+} }{ \text{BGT}_{\text{gs}} }$
\\
\hline
Krofcheck
&
${}^{71}\text{Ga} (p,n) {}^{71}\text{Ge}$
&
$< 0.057$
&
$
0.126
\pm
0.023
$
\\
Haxton
&
Shell Model
&
$
0.19
\pm
0.18
$
&
\\
Frekers
&
${}^{71}\text{Ga} ({}^{3}\text{He},{}^{3}\text{H}) {}^{71}\text{Ge}$
&
$
0.040
\pm
0.031
$
&
$
0.207
\pm
0.016
$
\\
JUN45
&
Shell Model
&
$
(
3.30
\pm
1.66
)
\times 10^{-2}
$
&
$
(
1.59
\pm
0.79
)
\times 10^{-2}
$
&
$
\left(
4.46
\pm
2.24
\right)
\times
10^{-6}
$
\\
\hline
\end{tabular}
}
\end{table}

\begin{table}[!tb]
\caption{
\label{tab:rat}
Ratios of measured and expected ${}^{71}\text{Ge}$ event rates
in the four radioactive source experiments, their correlated average,
and the statistical significance of the gallium anomaly
obtained with the cross sections in Table~\ref{tab:crs}.
}
\centering
{
\renewcommand{\arraystretch}{1.45}
\begin{tabular}{lcccccc}
\hline
&
GALLEX-1
&
GALLEX-2
&
SAGE-1
&
SAGE-2
&
Average
&
Anomaly
\\
\hline
$R_{\text{Bahcall}}$
&
$
0.95
\pm
0.11
$
&
$
0.81
\pm
0.11
$
&
$
0.95
\pm
0.12
$
&
$
0.79
\pm
0.08
$
&
$
0.85
\pm
0.06
$
&
$
2.6\sigma
$
\\
$R_{\text{Haxton}}$
&
$
0.86
\pm
0.13
$
&
$
0.74
\pm
0.12
$
&
$
0.86
\pm
0.14
$
&
$
0.72
\pm
0.10
$
&
$
0.76
\pm
0.10
$
&
$
2.5\sigma
$
\\
$R_{\text{Frekers}}$
&
$
0.93
\pm
0.11
$
&
$
0.79
\pm
0.11
$
&
$
0.93
\pm
0.12
$
&
$
0.77
\pm
0.08
$
&
$
0.84
\pm
0.05
$
&
$
3.0\sigma
$
\\
$R_{\text{JUN45}}$
&
$
0.97
\pm
0.11
$
&
$
0.83
\pm
0.11
$
&
$
0.97
\pm
0.12
$
&
$
0.81
\pm
0.08
$
&
$
0.88
\pm
0.05
$
&
$
2.3\sigma
$
\\
\hline
\end{tabular}
}
\end{table}

The gallium anomaly was originally discovered
\cite{Abdurashitov:2005tb,Laveder:2007zz,Giunti:2006bj,Acero:2007su}
using the Bahcall cross sections \cite{Bahcall:1997eg}
reported in the first line of Table~\ref{tab:crs},
that have been obtained using the
$\text{BGT}$'s measured in 1985 in the
$(p,n)$
experiment of Krofcheck et al. \cite{Krofcheck:1985fg,Krofcheck-PhD-1987}
(see Table I of Ref.~\cite{Bahcall:1997eg}),
listed in the first line of Table~\ref{tab:bgt}.
The cross sections of $^{51}$Cr and $^{37}$Ar electron neutrinos
can be calculated from the Gamow-Teller strengths through
\begin{equation}
\sigma
=
\sigma_{\text{gs}}
\left(
1
+
\xi_{5/2-}
\frac{\text{BGT}_{5/2-}}{\text{BGT}_{\text{gs}}}
+
\xi_{3/2-}
\frac{\text{BGT}_{3/2-}}{\text{BGT}_{\text{gs}}}
+
\xi_{5/2+}
\frac{\text{BGT}_{5/2+}}{\text{BGT}_{\text{gs}}}
\right)
,
\label{csbgt}
\end{equation}
with the phase-space coefficients \cite{Bahcall:1997eg}
\begin{align}
\null & \null
\xi_{5/2-}({}^{51}\text{Cr}) =0.663
\null & \null
\null & \null
\xi_{3/2-}({}^{51}\text{Cr}) =0.221
\null & \null
\null & \null
\xi_{5/2+}({}^{51}\text{Cr}) = 0,
\label{xiCr}
\\
\null & \null
\xi_{5/2-}({}^{37}\text{Ar}) =0.691
\null & \null
\null & \null
\xi_{3/2-}({}^{37}\text{Ar}) =0.262
\null & \null
\null & \null
\xi_{5/2+}({}^{37}\text{Ar}) =0.200
\label{xiAr}
\end{align}
and
\cite{Bahcall:1997eg}
\begin{align}
\sigma_{\text{gs}}({}^{51}\text{Cr})
=
\left(
5.53
\pm
0.01
\right) \times 10^{-45}  \, \text{cm}^2 
\,,
\label{gsCr}
\\
\sigma_{\text{gs}}({}^{37}\text{Ar})
=
\left(
6.62
\pm
0.01
\right) \times 10^{-45} \, \text{cm}^2
\,.
\label{gsAr}
\end{align}
The first line in Table~\ref{tab:rat}
gives the ratios of measured and expected ${}^{71}\text{Ge}$ event rates
in the four radioactive source experiments and their correlated average
obtained using the Bahcall cross section,
which led to a
$2.6\sigma$
gallium anomaly.

In 1998 Haxton \cite{Haxton:1998uc} published
a shell model calculation of $\text{BGT}_{5/2-}$
that gave the relatively large value in the second line of Table~\ref{tab:bgt},
albeit with a very large uncertainty.
The cross sections obtained with the Haxton $\text{BGT}_{5/2-}$
and the Krofcheck et al. measurement of $\text{BGT}_{3/2-}$
are listed in the second line of Table~\ref{tab:crs}.
As one can see from Table~\ref{tab:rat}
the larger uncertainties of the Haxton cross sections lead to a slight decrease
of the gallium anomaly
from the Bahcall
$2.6\sigma$
to
$2.5\sigma$,
in spite of the larger Haxton cross sections.

In 2011 Frekers et al.
\protect\cite{Frekers:2011zz}
published the measurements of
$\text{BGT}_{5/2-}$ and $\text{BGT}_{3/2-}$
in the third line of Table~\ref{tab:bgt},
obtained with
${}^{71}\text{Ga} ({}^{3}\text{He},{}^{3}\text{H}) {}^{71}\text{Ge}$
scattering.
They found a finite value of $\text{BGT}_{5/2-}$,
albeit with a large uncertainty,
which is compatible with the upper limit of
Krofcheck et al. \cite{Krofcheck:1985fg,Krofcheck-PhD-1987}.
On the other hand,
the Frekers et al. value of $\text{BGT}_{3/2-}$
is about $2.9\sigma$ larger than that of Krofcheck et al.
If one considers these Gamow-Teller strengths as applicable to the $\nu_{e}$--$^{71}\text{Ga}$
cross section without corrections due to the tensor contributions
(that would require a theoretical calculation),
there is a significant increase of the
${}^{51}\text{Cr}$ and ${}^{37}\text{Ar}$ neutrino
cross sections with respect to the Bahcall cross sections
and an increase of the gallium anomaly to $3.0\sigma$,
as shown in Table~\ref{tab:rat}.

From Table~\ref{tab:crs} one can also see that
our JUN45 shell-model calculation of the Gamow-Teller strengths,
listed in the fourth row of Table~\ref{tab:bgt},
gives cross sections that are smaller than the previous ones.
As a result,
the gallium anomaly decreases to $2.3\sigma$,
as shown in Table~\ref{tab:rat}.

\begin{figure}[t!]
\begin{minipage}[t]{0.49\textwidth}
\begin{center}
\includegraphics*[width=\linewidth]{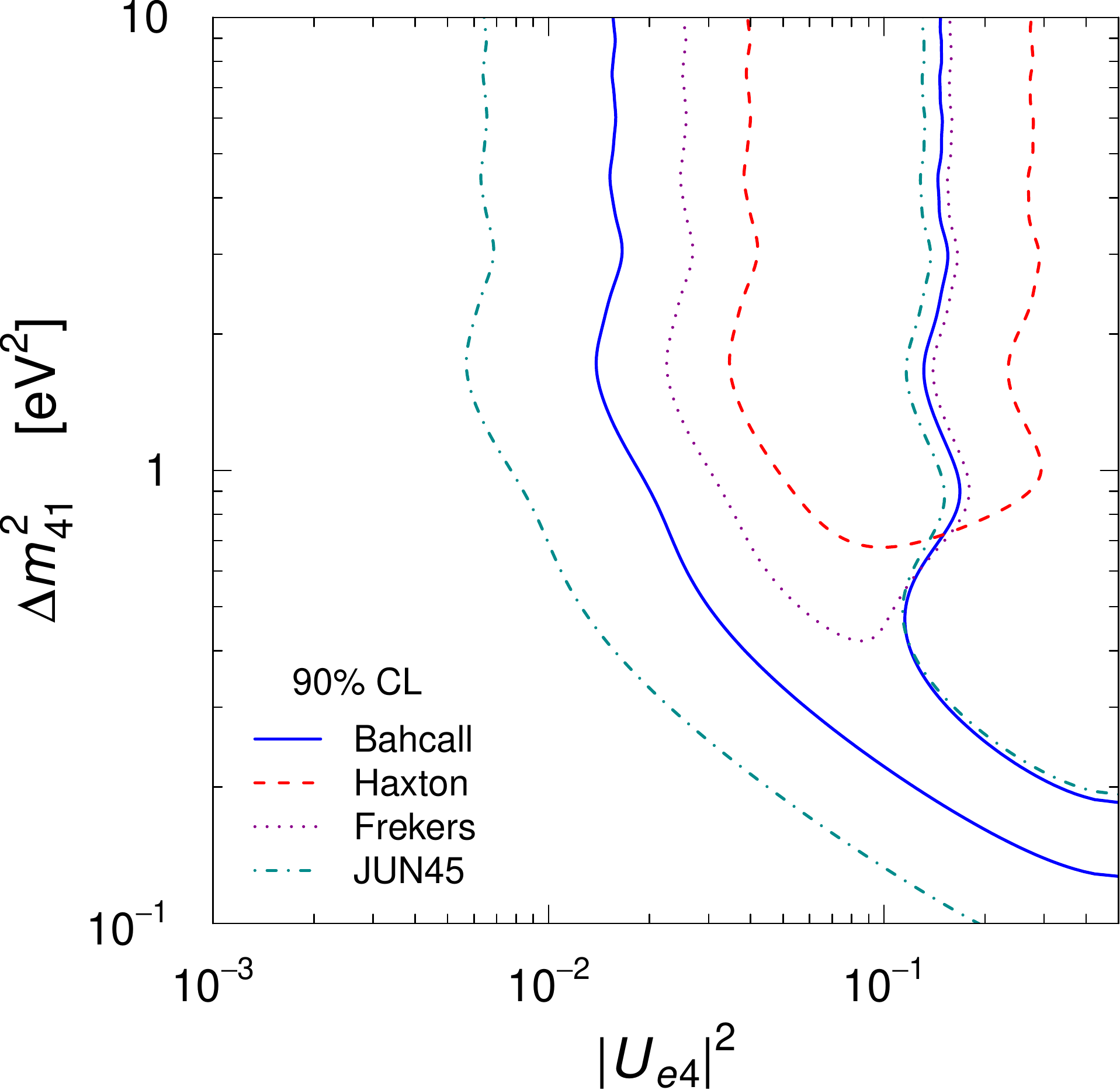}
\end{center}
\caption{ \label{fig:ue4-gal}
Comparison of the 90\% allowed regions
in the $|U_{e4}|^2$--$\Delta{m}^2_{41}$ plane
obtained with the cross sections in Table~\ref{tab:crs}.
The Bahcall and JUN45 allowed regions are between the two corresponding curves.
The Haxton and Frekers allowed regions are enclosed by the corresponding curves,
without an upper limit on $\Delta{m}^2_{41}$.
}
\end{minipage}
\hfill
\begin{minipage}[t]{0.49\textwidth}
\begin{center}
\includegraphics*[width=\linewidth]{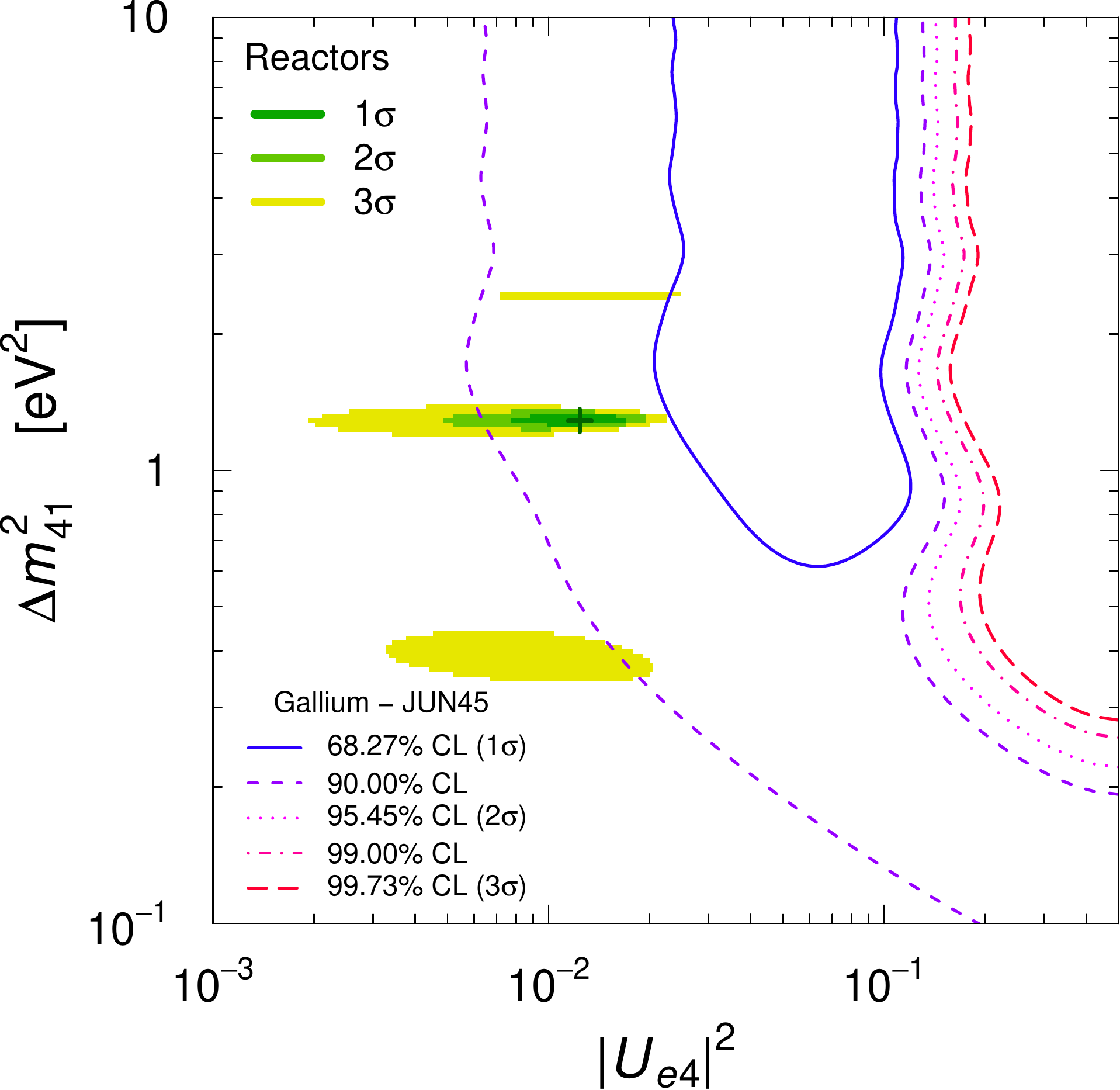}
\end{center}
\caption{ \label{fig:ue4-rea}
Comparison of the allowed regions
in the $|U_{e4}|^2$--$\Delta{m}^2_{41}$ plane
obtained from the Gallium data with the JUN45 cross sections
and
the allowed regions obtained from the analysis of the data of the
NEOS, DANSS and PROSPECT reactor experiments.
}
\end{minipage}
\end{figure}

The gallium anomaly has been considered as one of the indications in favor of
short-baseline neutrino oscillations due to active-sterile neutrino mixing
(see the reviews in Refs.~\cite{Gariazzo:2015rra,Giunti:2019aiy,Diaz:2019fwt,Boser:2019rta}).
In the framework of the 3+1 mixing scheme,
which is the simplest one that extends the standard three-neutrino mixing with the addition
of a sterile neutrino at the eV mass scale,
the survival probability of electron neutrinos and antineutrinos\footnote{
In general,
CPT invariance implies the equality of the survival probability of neutrinos and antineutrinos
of each flavor (see, for example, Ref.~\cite{Giunti:2007ry}).
}
in short-baseline experiments is given by
\begin{equation}
P_{\nua{e}\to\nua{e}}^{\text{SBL}}
=
1
-
4 |U_{e4}|^2 \left( 1 - |U_{e4}|^2 \right)
\sin^2\!\left( \dfrac{ \Delta{m}^2_{41} L }{ 4 E } \right)
,
\label{pee}
\end{equation}
where
$L$ is the source-detector distance,
$E$ is the neutrino energy,
$U$ is the unitary $4\times4$ neutrino mixing matrix,
and
$\Delta{m}^2_{41} = m_{4}^2 - m_{1}^2$ is the squared-mass difference between a
heavy, almost sterile, $\nu_{4}$ with mass $m_{4} \sim 1 \, \text{eV}$
and the standard three light neutrinos $\nu_{1}$, $\nu_{2}$, and $\nu_{3}$
with respective masses
$m_{1}$, $m_{2}$, and $m_{3}$
much smaller than $1 \, \text{eV}$
(hence, $\Delta{m}^2_{41} \simeq \Delta{m}^2_{42} \simeq \Delta{m}^2_{43}$
in Eq.~(\ref{pee})).

Figure~\ref{fig:ue4-gal}
shows the differences of the 90\% allowed regions
in the $|U_{e4}|^2$--$\Delta{m}^2_{41}$ plane
obtained from the gallium data with the four cross sections in Table~\ref{tab:crs}.
One can see that the Haxton and Frekers cross sections
give a relatively large gallium anomaly,
with preferred regions at
$0.03 \lesssim |U_{e4}|^2 \lesssim 0.2$
and
$\Delta{m}^2_{41} \gtrsim 0.5-0.7 \, \text{eV}^2$.
The Bahcall cross sections allow lower values of
$|U_{e4}|^2$ and $\Delta{m}^2_{41}$
and our JUN45 shell model calculation allows still lower values,
as low as $|U_{e4}|^2 \gtrsim 0.007$ for $\Delta{m}^2_{41} \gtrsim 1 \, \text{eV}^2$.

The indication in favor of short-baseline $\nu_{e}$ disappearance due to active-sterile mixing
is at the level of
1.9 
($\Delta\chi^2=5.7$ 
with 2 degrees of freedom with respect to the absence of oscillations).
This value must be compared with the
2.2, 
2.7, 
 and 2.6 
levels obtained with the
Bahcall, Haxton, and Frekers
cross sections, respectively.

It is also interesting to compare our results for the gallium anomaly
with the recent indication in favor of short-baseline electron neutrino and antineutrino
disappearance
\cite{Gariazzo:2018mwd,Dentler:2018sju}
obtained from the combined analysis of the data of the
NEOS \cite{Ko:2016owz}
and
DANSS \cite{Alekseev:2018efk}
reactor experiments.
This indication is independent of our knowledge of the reactor antineutrino fluxes,
because it is obtained from comparisons of the detection energy spectra at different distances
from the reactor source.
Hence, it depends only on the experimental uncertainties,
not on the theoretical uncertainties of the neutrino rates and spectra
that are widely considered to be larger than those estimated before the discovery of
the mysterious 5 MeV bump (see, for example, Ref.~\cite{Hayes:2016qnu}).

A comparison of the results for the gallium anomaly
obtained with our JUN45 shell-model calculation
with the NEOS and DANSS
indications in favor of short-baseline oscillations
is interesting because the comparison presented in Ref.~\cite{Gariazzo:2018mwd},
where the Frekers cross sections have been used,
indicated an incompatibility of the $2\sigma$ allowed regions,
with a tension quantified by a parameter goodness of fit of 4\%.

Figure~\ref{fig:ue4-rea}
shows the comparison of the allowed regions
in the $|U_{e4}|^2$--$\Delta{m}^2_{41}$ plane
obtained with our JUN45 shell model for different confidence levels
with the regions obtained 
from the combined analysis of the data of the
NEOS and DANNS reactor experiments,
to which we have added the more recent data of the
PROSPECT \cite{Ashenfelter:2018iov} reactor experiment
that excludes large values of $|U_{e4}|^2$ for
$0.7 \lesssim \Delta{m}^2_{41} \lesssim 7 \, \text{eV}^2$.
One can see that there is an overlap of the 90\% CL allowed regions,
indicating a reasonable agreement between the gallium anomaly and the
reactor data.
The corresponding parameter goodness of fit is a favorable
$16 \%$ 
($\Delta\chi^2/\text{NDF}
=
3.6
/
2).$

\section{Conclusions}

In this Letter we presented the results from large-scale shell-model calculations 
regarding the scattering of $^{37}\rm Ar$ and $^{51}\rm Cr$ neutrinos off the
$^{69,71}\rm Ga$ isotopes. The new theoretical estimates for 
these cross sections are $6.80\pm0.12\times 10^{-45}$ cm$^2$ and 
$5.67\pm0.10\times 10^{-45}$ cm$^2$ respectively which are 2.5--3.0\% lower 
than the previous predictions.

According to our JUN45 shell-model calculation of the cross sections of the interaction of
$\nu_{e}$'s produced by $^{51}$Cr and $^{37}$Ar radioactive sources with $^{71}$Ga,
the gallium anomaly related to the GALLEX and SAGE experiments
is weaker than that obtained in previous evaluations, decreasing the significance 
from 3.0$\sigma$ to 2.3$\sigma$. 
Our result is compatible with the recent indication in favor of short-baseline
$\bar\nu_{e}$ disappearance due to small active-sterile neutrino mixing
obtained from the combined analysis of the data of the
NEOS and DANSS
reactor experiments.

The possible sources for the difference between the new theoretical cross sections 
and those predicted by charge-exchange reactions were examined. It is pointed out that 
the cross section of the scattering to the 500 keV $3/2^-$ state in 
$^{71}\rm Ge$ is most likely overestimated in the charge-exchange reaction due to a particular 
combination of destructive and constructive interferences between Gamow-Teller 
and tensor contributions. 

\section*{Acknowledgements}
This work has been partially supported by the Academy of Finland under the Academy 
project no. 318043. J. K. acknowledges the financial support from Jenny and Antti
Wihuri Foundation. We would like to thank Prof. K. Zuber for his suggestion to 
tackle this topic and Prof. H. Ejiri for enlightening discussions.


\end{document}